\documentclass
[showpacs,superscriptaddress,prl,twocolumn,tightenlines,balancelastpage,10pt,a4paper]{revtex4}%
\usepackage{graphicx}% Include figure files
\usepackage{subfigure}% From CTAN/macros/latex/contrib/supported/subfigure
\usepackage{dcolumn}% Align table columns on decimal point
\usepackage{bm}% bold math

\begin{document}
\preprint{To be revised}
\title{Experimental Fault-Tolerant Quantum Cryptography in a Decoherence-Free Subspace}

\author{Qiang Zhang}
\affiliation{Hefei National Laboratory for Physical Sciences at
Microscale \& Department of Modern Physics, University of Science
and Technology of China, Hefei, Anhui 230026, P.R. China}
 \affiliation{Physikalisches Institut,
Universit\"{a}t Heidelberg, Philosophenweg 12, 69120 Heidelberg,
Germany}
\author{Juan Yin}
\affiliation{Hefei National Laboratory for Physical Sciences at
Microscale \& Department of Modern Physics, University of Science
and Technology of China, Hefei, Anhui 230026, P.R. China}
\author{Teng-Yun Chen}
\affiliation{Hefei National Laboratory for Physical Sciences at
Microscale \& Department of Modern Physics, University of Science
and Technology of China, Hefei, Anhui 230026, P.R. China}
\author{Shan Lu}
\affiliation{Hefei National Laboratory for Physical Sciences at
Microscale \& Department of Modern Physics, University of Science
and Technology of China, Hefei, Anhui 230026, P.R. China}
\author{Jun Zhang}
\affiliation{Hefei National Laboratory for Physical Sciences at
Microscale \& Department of Modern Physics, University of Science
and Technology of China, Hefei, Anhui 230026, P.R. China}
\author{Xiao-Qiang Li}
\affiliation{Hefei National Laboratory for Physical Sciences at
Microscale \& Department of Modern Physics, University of Science
and Technology of China, Hefei, Anhui 230026, P.R. China}
\author{Tao Yang}
\affiliation{Hefei National Laboratory for Physical Sciences at
Microscale \& Department of Modern Physics, University of Science
and Technology of China, Hefei, Anhui 230026, P.R. China}
\author{Xiang-Bin Wang}
\affiliation{Imai Quantum Computation
and Information Project, ERATO, JST, Daini Hongo White Building
201,5-28-3, Hongo, Bunkyo, Tokyo 113-0033, Japan}
\author{Jian-Wei Pan}
\affiliation{Hefei National Laboratory for Physical Sciences at
Microscale \& Department of Modern Physics, University of Science
and Technology of China, Hefei, Anhui 230026, P.R. China}
\affiliation{Physikalisches Institut, Universit\"{a}t Heidelberg,
Philosophenweg 12, 69120 Heidelberg, Germany}

\date{\today}

\begin{abstract}
We experimentally implement a fault-tolerant quantum key
distribution protocol with two photons in a decoherence-free
subspace (DFS). It is demonstrated that our protocol can yield
good key rate even with large bit-flip error rate caused by
collective rotation, while the usual realization of BB84 protocol
cannot produce any secure final key given the same channel. Since
the experiment is performed in polarization space and does not
need the calibration of reference frame, important applications in
free-space quantum communication are expected. Moreover, our
method can also be used to robustly transmit an arbitrary
two-level quantum state in a type of DFS.
\end{abstract}
 \maketitle

Quantum key distribution (QKD) can help two remote parties to
accomplish unconditionally secure communications which is an
impossible task by any classical method \cite{gisin}. The security
of QKD is guaranteed by known principles of quantum mechanics
\cite{wooters,ekert,shor} rather than the assumed computational
complexity in classical secure communication. Since the first QKD
protocol proposed by Bennett and Brassard in 1984 (BB84 protocol)
\cite{benneett}, much work has been done in the field. In recent
years, numerous modified protocols have been proposed and
experimentally realized, e.g. the single-photon realizations in
either phase-coding or polarizations, the realizations with
entangled photon pairs and so on \cite{gisin}.

While each protocols or the physical realizations may have its own
advantage, there are still some limitations for QKD in practice.
In certain cases, we have no way to use the optical fibers and the
task has to be done in free space, for example, if we want to
carry out secure communications between a fixed station on the
earth and a moving object such as an airplane or satellite in the
space \cite{free-space}. Photon polarization is a natural
candidate for the QKD in free space, but the communicating parties
must share a common reference frame for spatial orientation
\cite{bartlett,rudolph} so that they can prepare and measure the
photon polarization in the same reference frame. Sometimes, it
could well be the case that the two parties have a relative
instantaneous rotation, for example during the quantum key
distribution between a swinging airplane and the earth. Moreover,
in some other cases the channel may also rotate the photon
polarization. Consequently, the two parties will no more share the
same reference frame from a passive perspective. These practical
disadvantages could bring significant error rate to the protocol
if one uses the single-photon polarization as information carrier,
in some extreme cases no secure final key can be distilled.

One possible solution to the above problem is to utilize
multi-qubit entangled states in a decoherence-free subspace (DFS)
where all the states are immune to some kind rotation of reference
frame. According to the informatics, the rotation of the reference
system can be seen as a collective noise, that is, the random
unitary transformation to each qubit is identical. The idea of DFS
\cite{palma,kwiat,kempe} was proven to be very important in
quantum computation and quantum communication.

Very recently, several quantum communication protocols based on
DFS have been put forward. Bartlett et.al \cite{bartlett} and
Boileau et.al \cite{boileau1} utilize four photons as a logic
qubit to perform quantum key distribution. Yet, the four photon
entanglement source based on nowadays technology is too poor to be
used in long distance communication. Recently, some other
protocols \cite{walton, boileau2} have been also put forward where
only two photons are used. Two photon entanglement source can be
achieved by spontaneous parameter down conversion (SPDC) and it
can be bright enough for the mission of quantum key distribution.
However, these protocols demand collective measurement of the two
photons after the trip through the channel and this kind of
measurement demand that the photons interfere with each other.

Another two-photon protocol suggested by one of us \cite{wang} has
the following properties: While two photons are requested and the
scheme only needs local individual measurement. Although the
protocol has the drawback that it only applies to the collective
random rotation noise, such a situation can be found in many
realistic applications such as in free space quantum communication
and communication with swinging object. In this letter, we report
an experimental realization of such a protocol. It is demonstrated
that our experimental method can yield good key rate even with
large bit-flip error rate caused by collective rotation, while the
usual realization of BB84 protocol cannot produce any secure final
key given the same channel.

Our experiment exploits the following 4 encoded BB84 states
\cite{wang}:
\begin{eqnarray}
|\overline
H\rangle&=&|\phi^+\rangle_{12}=\frac{1}{\sqrt{2}}(|H\rangle_1|H\rangle_2+|V\rangle_1|V\rangle_2)\nonumber
\\
|\overline
V\rangle&=&|\psi^-\rangle_{12}=\frac{1}{\sqrt{2}}(|H\rangle_1|V\rangle_2-|V\rangle_1|H\rangle_2)
\\
|+'\rangle&=&\frac{1}{\sqrt{2}}(|\overline H\rangle+|\overline
V\rangle)=\frac{1}{\sqrt{2}}(|H\rangle_1|+\rangle_2-|V\rangle_1|-\rangle_2)\nonumber
\\
|-'\rangle&=&\frac{1}{\sqrt{2}}(|\overline H\rangle-|\overline
V\rangle)=\frac{1}{\sqrt{2}}(|H\rangle_1|-\rangle_2+|V\rangle_1|+\rangle_2)\nonumber.
\end{eqnarray}

Here $|H\rangle, |V\rangle, |+\rangle, |-\rangle$ are the same
meaning as in BB84 protocol, represent for horizontal, vertical,
and diagonal and anti-diagonal polarization states respectively.
It is easy to verify, the states $|\psi^-\rangle_{12}$ and
$|\phi^+\rangle_{12}$ are invariant under the following collective
rotation

\begin{eqnarray}
|H\rangle&\Rightarrow&\cos\theta|H\rangle-\sin\theta|V\rangle\nonumber\\
|V\rangle&\Rightarrow&\sin\theta|H\rangle+\cos\theta|V\rangle.
\end{eqnarray}

Here, $\theta$ is the collective rotation noise parameter, which
is depending on the environment and will fluctuate with time. This
invariance implies that all the linear superposition of the two
states constitute a subspace that is decoherence free to the
collective rotation noise.

\begin{figure}
\includegraphics[width=3in]{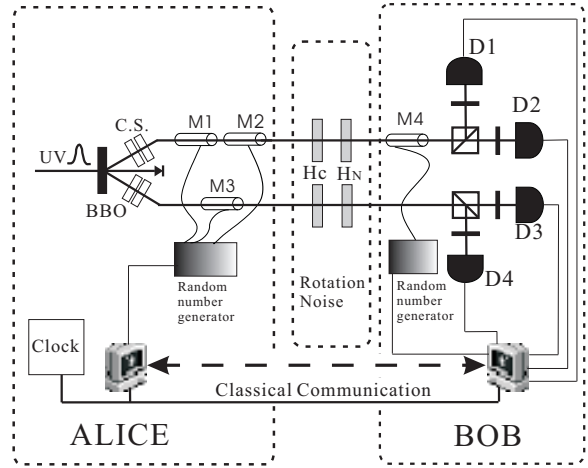}
 \caption{ \label{fig:figure1}Experimental setup for two photon
fault-tolerant quantum key distribution protocol. The $394$ nm UV
pulses are produced by frequency doubling the $788$ nm pulses of
the mode-locked laser using a nonlinear LBO crystal $(LiB_3O_5)$.
The UV pulses pass through 2mm BBO ($\beta$-barium borate )
crystal and polarization entangled photon pairs in the state,
$|\psi^-\rangle_{12}$, are created. In order to compensate the
birefringence of the BBO crystal, we place a half wave plate (HWP)
and a compensating BBO crystal of 1 mm thickness on each path of
the two photons as a compensating system (labelled C.S.). Three
electro-optic modulators controlled by a random number generator
are utilized to produce the other three two-photon states
requested by the protocol. They are labelled by M1-M3
respectively. Channel noise of random rotation is realized by four
half wave plates, two in each path ($H_C$ and $H_N$). We use an
electro-optic modulator M4 controlled by another random number
generator to choose the measurement bases. After the modulator, we
place polarized beam splitters (PBS) and the raw key is obtained
by observing the clicking of detectors behind the PBS. A
interference filters (IF) with $FWHM = 2.8$nm is placed before
each single photon detector (D1 - D4) to improve the visibility of
the entangled pair. Each user has a computer to control the random
number generator and record the detector's events.}
\end{figure}

The experimental setup of the protocol is sketched in
Fig.~\ref{fig:figure1}. Type II parametric down-conversion in
$\beta$-barium borate (BBO), pumped by a mode-locked femtosecond
laser working at wavelength of $394$nm and a power of $600$mW,
produces about 4000 polarization entangled photon pairs per second
at $788$nm whose state is $|\psi^-\rangle_{12}$ , i.e. the state
$|\overline V\rangle$ in our protocol. The other three states can
be obtained by performing a corresponding local unitary
transformation on the state $|\overline V\rangle$.

We use electro-optic modulators controlled by random number
generators to realize Alice's encryption. After the bias voltage
and half-wave voltage being carefully calibrated and adjusted, the
modulators can translate the photon's state properly. When the
modulators are turned off, they do nothing to the polarization of
the photons to be sent. Once switched on, the modulators will
change the polarization of the photons like half wave plates.
Modulator 1 is set to be $0$ degree to its axis, Modulator $2$ and
3 are set to be $45$ degree and $22.5$ degree, respectively. It is
easy to show that when modulator $1,2$ are turned on together, the
state will be changed from $|\overline V\rangle$ to $|\overline
H\rangle$. When modulator $1,3$ are turned on, the state is
$|+'\rangle$ and when modulator $2,3$ are turned on, $|-'\rangle$
is produced. Obviously when all the modulators are switched off,
the output state is $|\overline V\rangle$.

Similar to the realization of BB84 protocol, Alice has two random
number $X$, $Y$. $X$ is used to choose base and $Y$ is Alice's bit
value. Alice utilizes the two random number to control the
modulators to randomly prepare one of the four encoded states in
the DFS. If $X=0$, Alice will choose the base $\{|\overline
H\rangle, |\overline V\rangle\}$. When X=1, she will choose the
base $\{|+'\rangle, |-'\rangle\}$. If Y=1, Alice will prepare
$|\overline H\rangle$ or $|+'\rangle$. Otherwise, she will prepare
$|\overline V\rangle$ or $|-'\rangle$ . Table \ref{tab:table1}
describes the process in detail.

\begin{table}
\caption{\label{tab:table1}Summary of the process of encoding. $X$
denotes the base and $Y$ is the bit value. They are prepared by
the random number generator. The two number determine which
modulators will be turn on and which state is prepared.}
\begin{ruledtabular}
\begin{tabular}{cccccc}
$X$ &$Y$ &Modulator 1 &Modulator 2 &Modulator3 &State\\
\hline
0 & 0 & 0 & 0 & 0 & $|\overline V\rangle$\\
0 & 1 & 1 & 1 & 0 & $|\overline H\rangle$\\
1 & 0 & 0 & 1 & 1 & $|-'\rangle$\\
1 & 1 & 1 & 0 & 1 & $|+'\rangle$\\
\end{tabular}
\end{ruledtabular}
\end{table}

The two random numbers are achieved by quantum process of
splitting a beam of single photons similar as Jennewein et al. did
in their experiment \cite{jennewein}. At first, The two random
numbers are stored in a FIFO memory. Then they will be readout and
encoded according to table \ref{tab:table1} triggered by a
$100$kHz clock. In our experiment, the encoding frequency of
$100$kHz is so high that the probability of more than $1$ pair
appearing in the same encoding period is small enough to guarantee
the security of quantum key distribution.

We use two half-wave plates (HWPs) to simulate the collective
random rotation of the noise channel. Here, the unitary
transformation introduced by the HWPs is slightly different from
the noise of collective random rotations as assumed in the
original protocol. Instead of the unitary transformation of
Eq.(2), if we set the HWP at the angle $\frac{\theta}{2}$ to its
optical axis, the function is as follows:
\begin{eqnarray}
|H\rangle&\Rightarrow&\cos\theta|H\rangle-\sin\theta|V\rangle\nonumber\\
|V\rangle&\Rightarrow&-(\sin\theta|H\rangle+\cos\theta|V\rangle).
\end{eqnarray}
In order to realize the rotation noise as in Eq. (2), we further
insert an additional HWP ($H_C$), which is set at $0$ degree, in
front of the $H_N$ to correct the minus phase shift in each path.

Since the four encoded states as shown in Eq. (1) are invariant
under the collective rotation described above, Bob only needs to
use an electro-optic modulator to choose his measurement bases and
then let each photon respectively pass through a PBS to perform a
polarization measurement (see Fig.~\ref{fig:figure1}). In this
way, our protocol avoids the collective measurement which needs
the two-photon interference \cite{walton, boileau2}. The entangled
photon pairs are detected by fiber-coupled single photon
detectors. Bob uses another random number generator $Z$ to control
the electro-optic modulator that is set at $22.5$ degree. If
$Z=0$, he measures photon 1 in the $\{|H\rangle, |V\rangle\}$
basis. Otherwise he chooses the $\{|+\rangle, |-\rangle\}$ basis.
For photon 2, as there is no modulator, it is measured in
$\{|H\rangle, |V\rangle\}$ basis.

The photons are detected by silicon avalanche photon diodes. When
Bob find that photon 1(D1 or D2) and photon 2(D3 or D4) are
simultaneously detected in a coincidence window of $5$ns, he will
record it as a successful detection event. If D1 and D4 or D2 and
D3 fire simultaneously, he will record the bit as ``$0$''.
Otherwise he will record as ``$1$''.

The encoding clock can also give a timing signal in a measurement
turn. The computer on Bob's side registers all detection events as
time stamps together with measure base information and the
detection result.  After the key distribution, Bob will declare at
what time he get a detection event and his measurement base. And
Alice will tell Bob to discard those bits in wrong bases to
produce the raw key. Then, they can do error test and final key
distillation. As it has been shown in Ref \cite{wang}, the
protocol here can actually be regarded as BB84 protocol with
encoding and decoding. Therefore, we only need to check its
quantum bit error rate (QBER) after decoding for the security
issue, i.e., if the QBER after decoding is less than $11\%$, then
we conclude that we can distill some unconditionally secure final
key \cite{mayers,shor}.

\begin{figure}
\includegraphics[scale=0.5]{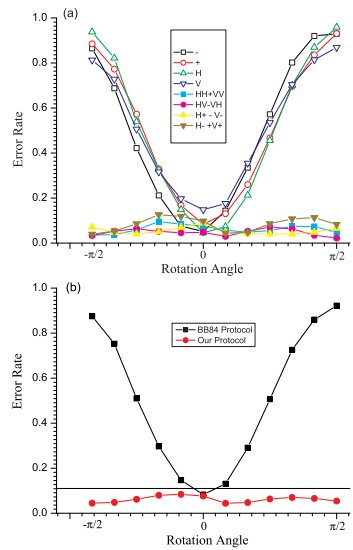}
\caption{\label{fig:figure2} (a)Experimentally result of the
quantum bit error rates of all the states in our protocol and BB84
protocol in the rotation noise. (b) Experimental result of the
total quantum bit error rate of our protocol and BB84 protocol. It
can be seen at any angle that our protocol is below the line of
$11\%$, which is the security bound of the BB84 type protocol. The
error rates of our protocol are due to the imperfection of the
entanglement resource, the detectors and the electro-optic
modulators. The error rates of BB84 protocol are sinusoidal and
are the same to our protocol at $0$ degree angle rotation noise,
which are in agreement with theoretical prediction.}
\end{figure}

Fig.~\ref{fig:figure2} provides QBERs of each state with the same
collective random rotation channel and the total error rates. In
our two-photon encoding experiment, the rate of the raw keys is
about 2000/s. Under different random rotation noise the QBERs are
all observed to be less than $11\%$, which is sufficient to
guarantee the absolute security of the protocol. For each
experimental point, we spend $50$ seconds to collect the raw keys
to measure the QBERs, which leads to an error bar of the QBERs of
$0.1\%$. Therefore our protocol indeed always works given whatever
unknown collective random rotation noise.

In our experiment, we want to see whether the protocol has
advantage to standard realization of BB84 therefore we only need
to compare the QBERs of two protocols with the same collective
random noise channel. Experimentally, we project photon 2 into the
state $|+\rangle$ as a trigger. Then photon 1 can be treated as a
single photon source to be in the state $|-\rangle$. We use
Modulator 1 and 2 to prepare the four encoded single-photon state
in the standard $BB84$ protocol. Modulator 4 and the PBS behind
are used to perform the necessary polarization measurement on
photon 1. The obtained QBERs under different rotation noise is
also shown in Fig.~\ref{fig:figure2}. The figure shows that as
long as $|\theta|\geq\pi/18$, the QBERs of the standard
realization of BB84 are larger than $11\%$, which consequently
leads to the failure of quantum key distribution
\cite{mayers,shor}.

\begin{figure}
\includegraphics[width=2.5in]{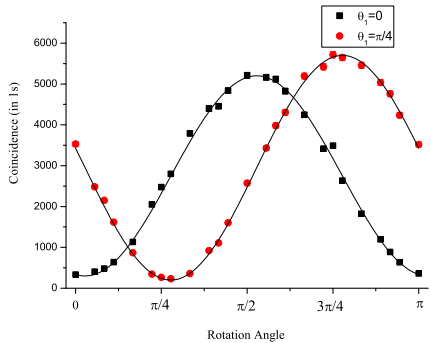}
\caption{\label{fig:figure3} Coincidence fringes for the entangled
photon source in our experiment whose state is
$|\psi^-\rangle_{12}$. When varying the polarizer angle
$\theta_1$, The two complimentary sine curve with a visibility of
$88\%$, which will bring $6\%$ error rate to the QKD protocol
which is the main reason of the QBER of our experiment.}
\end{figure}

It is important to note that, given perfect source of entangled
photon source, modulator and detector, the QBERs of our protocol
should approach to 0 at any random collective rotation noise.
However, in our two-photon quantum key distribution experiment a
significant average QBER of $6\%$ is observed. This is mainly due
to the imperfection of our entangle photon source from type II
parameter down conversion. As shown in Fig.~\ref{fig:figure3}, the
visibility of our entangled photon source has a limited visibility
of about $88\%$, which is in good agreement with our observed QBER
of $6\%$

In summary, we have experimentally realized a fault tolerant
quantum key distribution protocol in a DFS. As far as we have
known, this is the first result of two photon quantum cryptography
experiment that conquers the rotation noise with a decoherence
free subspace\footnote{ Recently, we become aware that a different
protocol for fault tolerant QKD has been realized by Jiang et al
\cite{jiang}.} We have verified the advantage of quantum key
distribution in DFS over a random collective rotation noise.

The experiment also has an extensive application background in
practice. Free space quantum communication is thought as a good
choice to realize global quantum communication \cite{free-space}.
In free space, the main noise is rotational type and the
dispersion noise can be neglected. Our protocol can be useful in
this situation. Also, in the cases when QKD between earth and
swinging objects is needed, our protocol has unique advantage. The
bit rate of our protocol is $2000$ per second and it can be
significantly improved by raising the laser power and improving
the detection efficiency. We believe it is potentially rather
useful for practical QKD in free space in the future.

Moreover, the experiment is completed in a DFS that plays an
important role in quantum computation and quantum communication.
As is known that there are two methods for robust quantum
communication, the quantum error correction codes(QECC) and the
decoherence free subspace. So far, QECC codes have not been
demonstrated by real qubits because they need at least 5 qubits.
Here we for the first time demonstrate robust quantum
communication of an arbitrary two-level quantum state in a type of
decoherence free subspace and we can transmit quantum information
robustly \cite{kempe}.

\begin{acknowledgments}
We thank Yu-Ao Chen for his useful help in picture. This work was
supported by the NNSF of China, the CAS, the PCSIRT and the
National Fundamental Research Program. This work was also
supported by the Marie Curie Excellent Grant of the EU, the
Alexander von Humboldt Foundation.
\end{acknowledgments}

\end{document}